\documentclass[%
 reprint,
 superscriptaddress,
 amsmath,amssymb,
 aps, 
]{revtex4-2}

\usepackage{graphicx}% Include figure files
\usepackage{dcolumn}% Align table columns on decimal point
\usepackage{bm}% bold math
\usepackage{multirow}
\usepackage{bm}% bold math
\usepackage{textcomp}
\usepackage{longtable}
\usepackage{booktabs}
\usepackage{epsfig}
\usepackage{rotating}
\usepackage{xfrac}
\usepackage{threeparttable}
\usepackage{lipsum} % generate dummy text for estimating page length
\usepackage{float}
\usepackage{dcolumn}
\usepackage{multirow}

\usepackage{rotating}
\RequirePackage{hyperref}
\hypersetup
{
unicode=false,%
pdftoolbar=true,
pdfmenubar=true,
pdffitwindow=false,
pdfstartview=FitH,
pdftitle={PRC},
pdfauthor={LAM Yi Hua},
pdfsubject={Nuclear Physics},
pdfcreator={LAM Yi Hua},
pdfproducer={LAM Yi Hua},
pdfkeywords={NucPhys},
bookmarksnumbered,%
colorlinks=true,%
linkcolor=magenta,%
citecolor=blue,% magenta
filecolor=cyan,%
urlcolor=blue,
breaklinks, % To break long figure/table caption in the List of Figures / Tables
plainpages=false}
\usepackage{CJKutf8}
\newcommand{\cntextSimKai}[1]{\begin{CJK*}{UTF8}{gkai}#1\end{CJK*}}
\newcommand{\cntextTraKai}[1]{\begin{CJK*}{UTF8}{bkai}#1\end{CJK*}}

\usepackage{xcolor}
\definecolor{Mycolor}{HTML}{C5E0B3}
\definecolor{red}{HTML}{ee6677}
\definecolor{blue}{HTML}{4477aa}
\definecolor{green}{HTML}{228833}
\definecolor{magenta}{HTML}{ee3377}
\definecolor{cyan}{HTML}{66ccee}
\definecolor{yellow}{HTML}{ccbb44}
\definecolor{grey}{HTML}{bbbbbb}
\begin{document}

\preprint{APS/123-QED}

\title{Nuclear shell evolution near $N=6$, 14, 20 and 28: insights from nuclear charge radii of short-lived nuclei derived from binding energies}%

\author{X.~Liu~(\cntextSimKai{刘星泉})}%
%\author{X.~Liu}%
\affiliation{Key Laboratory of Radiation Physics and Technology of the Ministry of Education, \href{https://ror.org/011ashp19}{Sichuan University}, Chengdu 610064,	China}

\author{W.~Chen~(\cntextSimKai{陈婉君})}%
%\author{W.~Chen}
\email{chenwanjun1106@outlook.com}
\affiliation{Key Laboratory of Radiation Physics and Technology of the Ministry of Education, \href{https://ror.org/011ashp19}{Sichuan University}, Chengdu 610064,	China}

\author{X.~Xu~(\cntextSimKai{徐星})}
\email{xuxing@impcas.ac.cn}
\affiliation{\href{https://ror.org/03x8rhq63}{Institute of Modern Physics}, Chinese Academy of Sciences, Lanzhou 730000, China}

\author{Yi~Hua~Lam~(\cntextTraKai{藍乙華})}%
%\author{Yi~Hua~Lam}%
\email{lamyihua@zstu.edu.cn}
%\homepage{https://lamyihua.github.io/}
\affiliation{Zhejiang Key Laboratory of Quantum State Control and Optical Field Manipulation, Department of Physics, \href{https://ror.org/03893we55}{Zhejiang Sci-Tech University}, 310018 Hangzhou, China}
\affiliation{Astrophysical Big Bang Laboratory, Pioneering Research Institute, \href{https://ror.org/01sjwvz98}{RIKEN}, Wako, Saitama 351-0198, Japan}

\author{H.~Zheng~(\cntextSimKai{郑华})}
\affiliation{School of Physics and Information Technology, \href{https://ror.org/0170z8493}{Shaanxi Normal University}, Xi'an 710119, China}

\author{J.~Han~(\cntextSimKai{韩纪锋})}
\affiliation{Key Laboratory of Radiation Physics and Technology of the Ministry of Education, \href{https://ror.org/011ashp19}{Sichuan University}, Chengdu 610064,	China}

\author{W.~Lin~(\cntextSimKai{林炜平})}
\affiliation{Key Laboratory of Radiation Physics and Technology of the Ministry of Education, \href{https://ror.org/011ashp19}{Sichuan University}, Chengdu 610064,	China}

\author{X.~Duan~(\cntextSimKai{段茜})}
\affiliation{Key Laboratory of Radiation Physics and Technology of the Ministry of Education, \href{https://ror.org/011ashp19}{Sichuan University}, Chengdu 610064,	China}

\author{X.~Zhang~(\cntextSimKai{张鑫})}
\affiliation{Key Laboratory of Radiation Physics and Technology of the Ministry of Education, \href{https://ror.org/011ashp19}{Sichuan University}, Chengdu 610064,	China}

\author{P.~Ren~(\cntextSimKai{任培培})}
\affiliation{Key Laboratory of Radiation Physics and Technology of the Ministry of Education, \href{https://ror.org/011ashp19}{Sichuan University}, Chengdu 610064,	China}

\date{\today}% It is always \today, today,

\begin{abstract}
	A deep understanding of the evolution of nuclear shell structure correlating with the nucleon number is crucial for unraveling the fundamental properties of the nuclear structure and for exploring new nuclear physics phenomena far from the $\beta$-stability line. Although significant progress has been made in probing nuclear shell evolution via the measurements of nuclear root-mean-square charge radii, $R_{\text{ch}}$, the scarcity of new data for short-lived and exotic nuclei due to the increasing difficulty of measurements presents a formidable challenge in obtaining deeper and more universal insights into the nature of shell evolution. To mitigate this issue, we develop an improved method, accounting for the exchange term, charge-symmetry breaking effect, and odd-even staggering effect in the Coulomb energy formulation compared with that proposed by Liu {\it et al.}~[\href{https://doi.org/10.1016/j.physletb.2025.140046}{Phys. Lett. B 872, 140046 (2026)}], to determine unmeasured $R_{\text{ch}}$ values. 
	Using the improved method, the $R_{\text{ch}}$ values of 59 nuclei are determined from their measured binding energies ($B$) and the respective $B$ and $R_{\text{ch}}$ of their mirror partners. We then systematically study the shell evolution near $N=6$, 14, 20 and 28 (sub)shells by placing the newly obtained $R_{\text{ch}}$ values into the corresponding isotopic chains. More comprehensive insights into the properties of nuclear shell evolution, particularly for the neutron-deficient sectors of the studied shell regions, e.g., $p$, $sd$ and $pf$ shells, are acquired, advancing our understanding of nuclear shell evolution in the light and intermediate mass region. 
\end{abstract}

\maketitle

%\tableofcontents

\section{\label{sec:intro}Introduction}

The atomic nucleus consisting of protons and neutrons exists in the bound state in a way that balancing between the attractive short-range strong nuclear force and the repulsive long-range Coulomb force. The bound state manifests the nuclear configurations and properties, e.g., mass and excitation spectra. Systematic investigations on the nuclei near the $\beta$-stability line have demonstrated that when the proton number, $Z$, or neutron number, $N$, corresponds to typical magic numbers, e.g., 2, 8, 20, 28, 50, 82, and 126, the nucleus, namely the magic nucleus, forms a highly stable closed-shell configuration. In 1949, Mayer \textit{et al.}~\cite{Mayer1949,Haxel1949} suggested the nuclear shell model that incorporates spin-orbit coupling interaction to reproduce the nuclear configuration of magic numbers of how nucleons populate the quantized single-nucleon energy levels. Since the 1990s, advances in radioactive ion beam technology \cite{Khasanov2025} have expanded the landscape of known nuclei from approximately 300 naturally stable and long-lived radioactive nuclei to over 3,000 identified nuclei. Investigations of the unstable nuclei under extreme isospin conditions with neutron excess or neutron deficiency have revealed a significant evolution of the nuclear shell structure. New findings indicate that the classic magic numbers aforementioned are either remarkably quenched or cease to exist, whereas new magic numbers and subshell structures appear in some specific regions. Understanding the evolution of the shell structure becomes one of the frontiers in exploring the nature of nuclear forces and extreme nuclear structure evolution \cite{Sorlin2008,Otsuka2020}. 

Among a variety of experimental observables, nuclear charge radius, $R_{\text{ch}}$, defined as the root-mean-square (rms) radius of the proton charge density distributions, manifests the unique characteristic of the shell evolution, as it is highly sensitive to the change of nuclear mean field and nucleon configuration mixing. The closure of the nuclear shell induces a relative contraction of the nuclear charge distribution and thus produces a local minimum of $R_{\text{ch}}$ in the magic nucleus compared to its neighboring nuclei \cite{Sorlin2008,Otsuka2020}, indicating a pronounce signature of the shell evolution. For instance, the recently measured $R_{\text{ch}}$ values of the neutron-deficient $^{40,41}\text{Sc}$ from collinear laser spectroscopy experiments at MSU show a surprising kink structure near the $N =20$ shell in contrast to its absence in the neighboring Ca, K, and Ar isotopic chains, demonstrating the unique shell evolution in the Sc isotopes relative to their neighboring nuclei~\cite{Konig2023}. By measuring the $R_{\text{ch}}$ and ground-state electromagnetic moments of indium isotopes as $N$ approaches 50 using laser spectroscopy, Karthein \textit{et al.}~found that $^{100}$Sn exhibits a doubly magic structure~\cite{Karthein2024}. Other cases were found in the $R_{\text{ch}}$ measurements for the $^{104-134}$Sn isotopes using two different collinear laser spectroscopy techniques at ISOLDE-CERN~\cite{Gustafsson2025}, where the measurements clarified the parabolic trend in $R_{\text{ch}}$ along the Sn isotopic chain between the $N=50$ and $N=82$ shell closures. These findings along with others~\cite{Sorlin2008,Otsuka2020} suggest the systematic variation of $R_{\text{ch}}$ can directly reflect the change in shell structure and provide a valuable experimental basis for investigating the nuclear-shell evolution.

Since the first measurement using high-energy elastic electron scattering by Rutherford~\cite{Rutherford1911}, the $R_{\text{ch}}$ measurements for around 1000 nuclei~\cite{Angeli2004,Wang2021_1} have been conducted using various techniques, i.e., muonic atom x-rays~\cite{Bazzi2011}, laser spectroscopy~\cite{Ewald2004}, optical and $K_{\alpha}$ x-ray isotope shifts~\cite{Angeli2004_1}, and the synergy of multiple techniques~\cite{Koszor2021}. Experimental measurements of $R_{\text{ch}}$ for most of the remaining short-lived and exotic nuclei have proven increasingly difficult due to their extremely low productions in the form of radioactive ion beams~\cite{Geldhof2022, Goodacre2021, Malbrunot2022, Pineda2021, Groote2020, Koszorus2021, YangXF2023}. Although significant progress has been made in investigating nuclear shell evolution from the perspective of $R_{\text{ch}}$, the lack of reliable data, especially for neutron-deficient nuclei, presents a formidable challenge in gaining deeper and more universal insights into the properties of nuclear shell evolution.

Comparing AME2020~\cite{mass2021_1} and the $R_{\text{ch}}$ databases~\cite{Angeli2004,Wang2021_1} indicates that for around 2500 nuclei, the number of nuclei measured with binding energies $B$ in high precision is more than twice the number of nuclei precisely measured with $R_{\text{ch}}$. The precisely measured $B$ are obtained from storage rings~\cite{Steck2020}, Penning traps~\cite{Dilling2018}, and beamline time-of-flight techniques~\cite{Matos2009}. Recently, based on the relation between nuclear binding energy difference, $\Delta B$, and $R_{\text{ch}}$ of mirror pairs, \citet{Liu2026} have developed a novel approach to determine the $R_{\text{ch}}$ of short-lived nuclei beyond the scope of available experimental measurements.  Using this method, we reported the first determination of the $R_{\text{ch}}$ values for ten neutron-deficient nuclei $^{26}\text{P}$, $^{27}\text{S}$, $^{28}\text{S}$, $^{41}\text{Ti}$, $^{43}\text{V}$, $^{45}\text{Cr}$, $^{46}\text{Cr}$, $^{47}\text{Mn}$, $^{50}\text{Fe}$ and $^{53}\text{Ni}$ from their newly measured masses with the advanced B$\rho$-defined isochronous mass spectrometry at the experimental cooler storage ring (CSRe) in Lanzhou~\cite{Zhang2023,Yu2024}. Such important inputs and method supplement the needed but missing important data and facilitate the systematic investigation of nuclear shell evolution via the $R_{\text{ch}}$ analysis.

According to the work of \citet{Liu2026}, a simple parameterization of the Coulomb energy, namely the Coulomb direct term, was adopted (the detailed description of this method is given in this paper). It is well known that the exchange term, which accounts for the quantum-mechanical exchange effect originating from the antisymmetrization of the wave function, contributes to $\Delta B$. In the 1960s, Okamoto, Nolen, and Schiffer found that theoretical calculations of Coulomb displacement energy remained systematically  lower than experimental values, even after incorporating various corrections, including finite nucleon size, vacuum polarisation, and core polarisation~\cite{Okamoto1964,Nolen1969}. This long-standing discrepancy is known as the ONS anomaly, and it was soon proposed that the missing contribution likely arises from a charge-symmetry breaking (CSB) effect of the strong nuclear force~\cite{Dong2018,Dong2019,Dong2020,Sagawa2024,Tanimura2025,Sun2025PLB}. In addition, the Coulomb energy exhibits odd-even staggering (OES), reflecting the short-range attractive pairing correlation among protons~\cite{Feenberg1946}. This indicates the necessity of including the staggering contribution in $\Delta B$.

This work aims to develop an improved method for determining the $R_{\text{ch}}$ values of unmeasured short-lived nuclei, based on the \citet{Liu2026} work. Methodologically, to further improve the predictive performance of $R_{\text{ch}}$ via the $\Delta B$-$R_{\text{ch}}$ relation, we refine the previously established method~\cite{Liu2026} by incorporating the contributions from the exchange, CSB, and OES effects into the $\Delta B$-$R_{\text{ch}}$ relation in this work. To validate the improvement of the present method, the $R_{\text{ch}}$ values predicted by our improved approach are benchmarked against those from the Bayesian neural network. The obtained $R_{\text{ch}}$ values from the improved method, mainly in the light and intermediate mass region, are then placed on their respective isotopic chains for a systematic shell evolution study. As demonstrated, the newly obtained $R_{\text{ch}}$ values shed light on the understanding of the shell properties of the $N=6$~\cite{Goeppert1963,Otsuka2001,Tran2018,Kanungo2002}, 14~\cite{Kanungo2002,Becheva2006}, 20~\cite{Sorlin2008,Otsuka2020}, and 28~\cite{Sorlin2008,Otsuka2020} (sub)shells that have attracted wide attention.

The paper is organized as follows. We briefly describe the improvement of the $R_{\text{ch}}$ determination method by incorporating the contributions from exchange effect, CSB effect, and OES effect in Sec.~\ref{sec:method}. Sections~\ref{sec:results} and \ref{sec:evolution} illustrate the determination of $R_{\text{ch}}$ values for unmeasured short-lived nuclei using the improved method. The evolution of shell structure at $N=6$, 14, 20, and 28 are discussed with the obtained $R_{\text{ch}}$ results subsequently. We then summarize the findings and present the perspectives in Sec.~\ref{sec:summary}.

\section{Principal improvement in determining $R_{\text{ch}}$ from $B$ and $R_{\text{ch}}$ for mirror nuclei}
\label{sec:method}

Historically, a semi-classical approach has been adopted 
in the earliest calculations of Coulomb energies, $E_{\text{C}}$, for nuclei~\cite{Shlomo1978}. The Coulomb energy of a nucleus with charge number $Z$ and a charge density distribution $\rho(\textbf{r})$ is given by
\begin{eqnarray}
	E_{\text{C}}= \displaystyle \int \rho(\textbf{r})U_{\text{C}}(\textbf{r})\text{d}^3\textbf{r} \propto \frac{Z^2}{R_{\text{real}}},
	\label{eq:eq1}
\end{eqnarray}
where $U_{\text{C}}(\textbf{r})$ is the Coulomb potential and $R_{\text{real}}$ is the real nuclear charge radius. Given that $R_{\text{real}}$ is  proportional to the cube root of mass number $A$ under the assumption that protons are uniformly distributed over the nucleus, the simple expression of Eq.~(\ref{eq:eq1}), namely direct Coulomb term, is used in the empirical mass formula~\cite{Weizs1935,Bethe1936} as 
\begin{eqnarray}
	E_{\text{C}}= -\displaystyle a_{\text{C}} \frac{Z^2}{A^{1/3}},
	\label{eq:eq2}
\end{eqnarray}
where $a_{\text{C}}$ is the Coulomb coefficient which can be extracted from the global fitting of experimental nuclear masses. The preceding minus sign indicates that the decrease in mass (or $B$) is attributed to Coulomb repulsion in given nuclei.

Bethe and Bacher found that Eq.~(\ref{eq:eq2}) does not consider the quantum-mechanical exchange effect which arises from the anti-symmetrization of the wave function~\cite{Bethe1936}. In the finite-range droplet model, the correction for the quantum-mechanical exchange considers the addition of the $Z^{4/3}$-dependent exchange term in Eq.~(\ref{eq:eq2})
\begin{eqnarray}
	E_{\text{C}}= -\displaystyle a_{\text{C}} \frac{Z^2}{A^{1/3}}\left[1-\frac{5}{4}\left(\frac{3}{2\pi}\right)^{2/3}Z^{-2/3} \right].
	\label{eq:eq4}
\end{eqnarray}
Equation (\ref{eq:eq4}) is one of the most widely used formulations for the Coulomb energy with the exchange term in the literature~\cite{Wangning2014}. If dropping the assumption of $R_{\text{real}}\propto A^{1/3}$ but adopting the more general proportional relationship between $R_{\text{real}}$ and $R_{\text{ch}}$, $ R_{\text{real}}\propto R_{\text{ch}}$, Eq.~(\ref{eq:eq4}) can be rewritten as 
\begin{eqnarray}
	E_{\text{C}}= -\displaystyle \tilde{a}_{\text{C}} \frac{Z^2}{R_{\text{ch}}}\left[1-\frac{5}{4}\left(\frac{3}{2\pi}\right)^{2/3}Z^{-2/3} \right] \, ,
	\label{eq:eq7}
\end{eqnarray}
of which the Coulomb coefficient correlated with $R_\mathrm{ch}$ is denoted as $\tilde{a}_{\text{C}}$. Then, based on the empirical mass formula~\cite{Gjorgievska2024}, the binding energy difference between a given mirror pair with interchanged numbers of $Z$ and $N$ ($Z>N$) can be derived as $\Delta B(Z,N) \equiv B(Z,N) -  B(N,Z) = E_{\text{C}}(Z,N)-E_{\text{C}}(N,Z) = \Delta E_{\text{C}}(Z,N)$, where $\Delta E_{\text{C}}(Z,N)$ is written as
\begin{align}
	\Delta E_{\text{C}}(Z,N)&= -\tilde{a}_{\text{C}} \left\{ \frac{Z^2}{R_{\text{ch}}(Z,N)} \left[ 1 - \frac{5}{4} \left( \frac{3}{2\pi} \right)^{\!2/3} Z^{-2/3} \right] \right. \notag \\
	&\hspace{1em} \!\left. - \frac{N^2}{R_{\text{ch}}(N,Z)} \left[ 1 - \frac{5}{4} \left( \frac{3}{2\pi} \right)^{\!2/3} N^{-2/3} \right] \right\}. 
	\label{eq:eq10}
\end{align}

The other improvement in the present work is to incorporate the CSB and OES effects into $\Delta B(Z,N)$. By accounting for these effects, $\Delta B(Z,N)$ can be further decomposed into three components as 
\begin{eqnarray}
	\Delta B(Z,N)= \Delta E_{\text{C}}(Z,N) + \Delta E_{\text{C}}^{\text{(CSB)}} + \Delta E_{\text{C}}^{\text{(OES)}},
	\label{eq:eq11}
\end{eqnarray}
where the terms, $\Delta E_{\text{C}}^{\text{(CSB)}}$ and $\Delta E_{\text{C}}^{\text{(OES)}}$, are the components originating from the charge symmetry breaking and odd-even staggering, respectively. In fact, $\Delta B(Z,N)$ correlates with the isovector coefficient of isobaric multiplet mass equation \cite{Dong2019, Lam2013b}, consisting of the dominant $\Delta E_{\text{C}}(Z,N)$ with contribution of the order of several MeVs.

In mean field approach, for the nucleus $(Z,N)$ taken as the minuend in $\Delta B(Z,N)$ of a given mirror pair, $\Delta E_{\text{C}}^{\text{(CSB)}}$ \cite{Dong2018,Dong2019,Dong2020} is given by 
\begin{eqnarray}
	\Delta E_{\text{C}}^{\text{(CSB)}}= +4Ta_{\text{sym}}^{\text{(CSB)}}(A,T_{z}) \, ,
	\label{eq:eq12}
\end{eqnarray} 
where $a_{\text{sym}}^{\text{(CSB)}}(A,T_{z})$ is the first-order symmetry energy coefficient for quantifying the CSB component, which can be determined via density functional integrals~\cite{Dong2018}, $T$ and $T_z$ denote the total isospin and its third component, respectively. The positive sign of the left-hand side corresponds to the present case where the proton-rich nucleus is $(Z>N)$ for the minuend. As shown by \citet{Dong2018}, the magnitude of $\Delta E_{\text{C}}^{\text{(CSB)}}$ depends on the physical assumptions in different models, leading to its inherent model dependence. Nevertheless, the deviations in the magnitude of $E_{\text{C}}^{\text{(CSB)}}$ are not pronounced across different models~\cite{Dong2018,Dong2019,Dong2020,Sagawa2024,Tanimura2025,Sun2025PLB}, and in Eq.~(\ref{eq:eq11}) its contribution is one to two orders of magnitude smaller than that of $\Delta E_{\text{C}}$. The $\Delta E_{\text{C}}^{\text{(CSB)}}$ merely contributes around 3~\% to $\Delta B(Z, N)$ when adopting the $a_{\text{sym}}^{\text{(CSB)}}(A,T_{z})$ values determined within the Brueckner-Hartree-Fock model without empirical parameter tuning~\cite{Dong2018,Dong2020}. Therefore, the obtained $a_{\text{sym}}^{\text{(CSB)}}(A,T_{z})$ values are used for the present work. Detailed validation of the principle for $a_{\text{sym}}^{\text{(CSB)}}(A,T_{z})$ can be found in Refs.~\cite{Dong2018,Dong2020}. Alternative methods for determining the CSB component are also available in Refs.~\cite{Lam2013a,RocaMaza2018,Naito2022,Naito2022_1,Naito2022_2,Novario2023,Sagawa2024,Tanimura2025,Sun2025PLB}. Certainly, the investigations of $\Delta E_{\text{C}}^{\text{(CSB)}}$ are important to reveal the detailed connection with nuclear origin.

We adopt a recent phenomenological formulation of Zong \textit{et al.}~\cite{Zong2022} for describing the $\Delta E_{\text{C}}^{\text{(OES)}}$ term, 
\begin{eqnarray}
	\Delta E_{\text{C}}^{\text{(OES)}} = -\frac{1-(-1)^{|Z-N|}}{2}(-1)^{Z}\frac{\beta}{A} \, ,
	\label{eq:eq13}
\end{eqnarray}
where $\beta$ has been optimized to be $1.473~(150)$~MeV based on AME2020~\cite{Zong2022}. The $\Delta E_{\text{C}}^{\text{(OES)}}$ only contributes less than 0.3~\% to $\Delta B(Z, N)$, one order of magnitude smaller than that of $\Delta E_{\text{C}}^{\text{(CSB)}}$. Substituting Eqs.~(\ref{eq:eq10}), (\ref{eq:eq12}) and (\ref{eq:eq13}) into Eq.~(\ref{eq:eq11}), we can establish the improved $\Delta B$-$R_{\text{ch}}$ relation as 
\begin{align}
	\Delta B(Z,N)&= -\tilde{a}_{\text{C}} \left\{ \frac{Z^2}{R_{\text{ch}}(Z,N)} \left[ 1 - \frac{5}{4} \left( \frac{3}{2\pi} \right)^{\!2/3} Z^{-2/3} \right] \right. \notag \\
	&\hspace{1em} \!\left. - \frac{N^2}{R_{\text{ch}}(N,Z)} \left[ 1 - \frac{5}{4} \left( \frac{3}{2\pi} \right)^{\!2/3} N^{-2/3} \right] \right\}. \notag \\
	&\hspace{1em} +4Ta_{\text{sym}}^{\text{(CSB)}}(A,T_{z}) \notag \\
	&\hspace{1em}-\frac{1-(-1)^{|Z-N|}}{2}(-1)^{Z}\frac{\beta}{A}.
	\label{eq:eq14}
\end{align}
Given the predetermined Coulomb coefficient $\tilde{a}_{\mathrm{C}}$ for the equation,  the $\Delta B$-$R_{\mathrm{ch}}$ relation offer pathways to determine the $R_{\text{ch}}$ values of the unmeasured short-lived nuclei, using the existing $B$ and $R_{\text{ch}}$ databases.

In formulating Eq.~(\ref{eq:eq14}), several other minor contributions, e.g., the Coulomb self-energy~\cite{Zong2022}, charge-independent breaking effect~\cite{Dong2018,Dong2019}, and Thomas-Ehrman shift \cite{Ehrman1951,Thomas1952} are neglected in the present work, due to their negligible higher-order corrections to $\Delta B(Z,N)$. Moreover, we only consider proton-rich nuclei away from the proton drip line. The ground state of these nuclei is sufficiently bound and dominated by higher-$\ell$ orbitals, making the Thomas-Ehrman shift negligible compared with the $\Delta E_\mathrm{C}(Z,N)$ energy, which is dominant in $\Delta B(Z, N)$. In Sec.~\ref{sec:improved_method}, we demonstrate that by incorporating the exchange, CSB, and OES effects, the improved method provides a more reliable approach for determining the $R_{\text{ch}}$ values of unmeasured short-lived nuclei, compared with the method based on Eq.~(\ref{eq:eq2})~\cite{Liu2026}.

\section{Results of $R_{\text{ch}}$ determination}
\label{sec:results}

\subsection{$\tilde{a}_{\text{C}}$ extraction}
\label{sec:ac_extration}

To determine the $R_{\text{ch}}$ for unmeasured short-lived nuclei from the existing $B$ and $R_{\text{ch}}$ data using the $\Delta B$-$R_{\text{ch}}$ relations for mirror nuclei, one of the most crucial steps is the extraction of $\tilde{a}_{\text{C}}$. Nevertheless, up to 2021, the available $B$ and $R_{\text{ch}}$ data is limited to only seven $|N-Z|\geq2$ mirror pairs, making universal properties of $\tilde{a}_{\text{C}}$ hard to be derived~\cite{Liu2026}. Although two additional mirror pairs of $^{32}\text{Ar}/^{32}\text{Si}$~\cite{Konig2024} and $^{40}\text{Sc}/^{40}\text{K}$~\cite{Konig2023} were recently supplemented, the $R_{\text{ch}}$ data from mirror pairs remains insufficient. To address the scarcity of mirror pair $R_{\text{ch}}$ data required for the $\tilde{a}_{\text{c}}$ extraction, a Bayesian neural network (BNN) method~\cite{Chen2025,Chen2025_1}, explicitly accounting for the uncertainties of $R_{\text{ch}}$ from the measurements and the uncertainties inherent in the model parametrizations, was used to supplement the $R_{\text{ch}}$ data. See the key treatments for BNN training and validation, and the comprehensive consideration of errors related to model parametrization and experimental uncertainties in Ref.~\cite{Liu2026}, which are critical for the extraction of $\tilde{a_{\text{C}}}$. 

\begin{figure}[tp]
	\centering
	\hspace*{0.3cm}
	\includegraphics[scale=0.4]{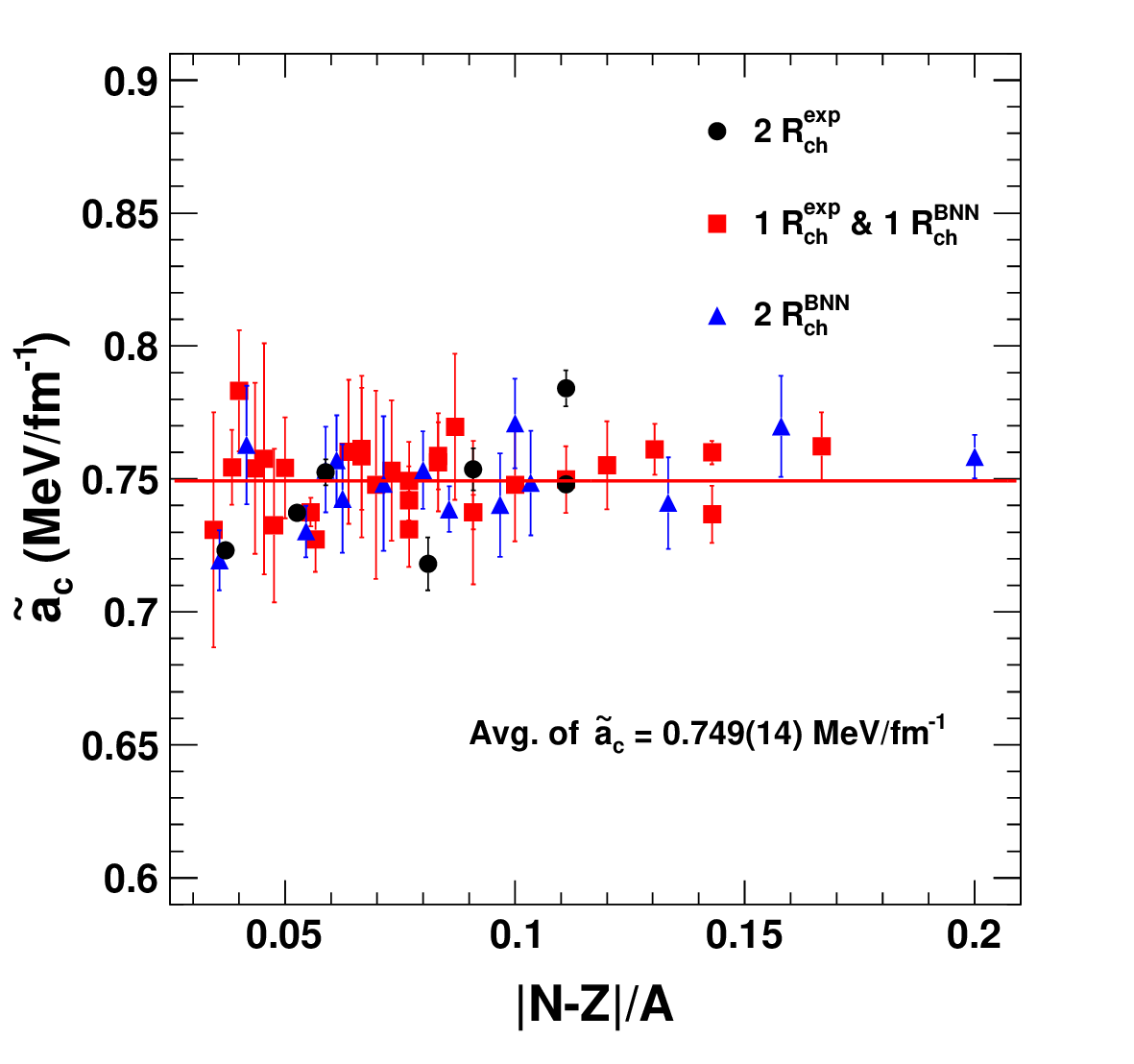}
	\caption{\footnotesize
		Coulomb coefficient $\tilde{a}_{\text{c}}$	as a function of $|N-Z|/A$. The $\tilde{a}_{\text{c}}$ are obtained from experimental $R_{\text{ch}}$ alone (black dots labeled by ``$2R_{\text{ch}}^{\text{exp}}$"), or from $R_{\text{ch}}^{\text{exp}}$ and experimental and BNN-predicted $R_{\text{ch}}$ (red squares labeled by ``$ 1 R_{\text{ch}}^{\text{exp}}\& 1 R_{\text{ch}}^{\text{BNN}}$"), or from BNN-predicted $R_{\text{ch}}$ alone (blue triangles labeled by ``$2R_{\text{ch}}^{\text{BNN}}$"). The solid line is the constant fit to the data points for guiding the eyes, and the average $\tilde{a}_{\text{c}}$ value is given with its uncertainty taken as the standard deviation. 
	}
	\label{fig:fig1}
\end{figure}

In the previous work \cite{Liu2026}, mirror pairs with $Z\geq 8$ and $|N-Z|\geq2$ are taken from the nuclei with available binding energies~\cite{Wang2021_1}. Then, the $\tilde{a}_{\text{C}}$ values of the identified mirror pairs, of which have one or two $R_{\text{ch}}$ values that are not measured, are deduced by substituting the missing data with the predicted $R_{\text{ch}}$ values. For this work, the same predicted $R_{\text{ch}}$ data by the BNN method are adopted~\cite{Liu2026}, and by following the same procedure, the $\tilde{a}_{\text{C}}$ values of the improved $\Delta B$-$R_{\text{ch}}$ relation of Eq.~(\ref{eq:eq14}) are deduced. The resultant $\tilde{a}_{\text{C}}$ values are plotted as a function of $|N-Z|/A$ in Fig.~\ref{fig:fig1}. We observe that every individual $\tilde{a}_{\text{c}}$ value for $2R_{\text{ch}}^{\text{exp}}$, $1 R_{\text{ch}}^{\text{exp}} \& 1 R_{\text{ch}}^{\text{BNN}}$, and $2R_{\text{ch}}^{\text{BNN}}$ is independent of $|N-Z|/A$ over the present interval of $|N-Z|/A\leq 0.2$. Similar $\tilde{a}_{\text{C}}$ behavior is also observed in the previous work~\cite{Liu2026}. Such numerical consistency of individual $\tilde{a}_{\text{C}}$ demonstrates the feasibility of deducing $\tilde{a}_{\text{C}}$ with the aid of the BNN extrapolating prediction. By averaging the combined $\tilde{a}_{\text{C}}$ data points from the experimental and BNN-predicted $R_{\text{ch}}$ values, the average $\tilde{a}_{\text{C}}$ values of the improved $\Delta B$-$R_{\text{ch}}$ relation is deduced to be  0.749~(14)~MeV/fm$^{-1}$ with the uncertainty to be the standard deviation, differing from that of the previous work, 0.714~(14)~MeV/fm$^{-1}$. This difference can be attributed to the incorporation of the exchange, CSB, and OES effects into the improved $\Delta B$-$R_{\text{ch}}$ relation in this work.

It is worth mentioning that the present $\tilde{a}_{\text{C}}$ extraction with the aid of the BNN learning does not require physical inputs about effective interactions or energy density functionals. Furthermore, the model-dependent contributions from the neural network parametrizations can be  suppressed to some extent by incorporating all available $R_{\text{ch}}$ predictions from the validated neural networks~\cite{Chen2025,Chen2025_1} into the $\tilde{a}_{\text{C}}$ extraction~\cite{Chen2025,Chen2025_1}.  The obtained  average $\tilde{a}_{\text{C}}$ from the combination of experimental and BNN-predicted $R_{\text{ch}}$ data is quasi-experimental, and can be utilized to determine the $R_{\text{ch}}$ values of the unmeasured nuclei.

\subsection{Improved method validation}
\label{sec:improved_method}

To ensure the performance of  $R_{\text{ch}}$ determination using the improved $\Delta B$-$R_{\text{ch}}$ relation, we first evaluate the $R_{\text{ch}}$ values for the ten neutron-deficient nuclei, $^{26}\text{P}$, $^{27}\text{S}$, $^{28}\text{S}$, $^{41}\text{Ti}$, $^{43}\text{V}$, $^{45}\text{Cr}$, $^{46}\text{Cr}$, $^{47}\text{Mn}$, $^{50}\text{Fe}$, and $^{53}\text{Ni}$, which have been determined in the previous work~\cite{Liu2026}. The results based on Eq.~(\ref{eq:eq14}) are listed in the second column of Table~\ref{table:table0}. For the benchmark validation, as the generalization capability of the BNN model for unmeasured nuclei has been carefully validated in the previous work~\cite{Chen2025,Chen2025_1}, the BNN-predicted values are adopted as the reference baseline and listed in the third column of Table~\ref{table:table0}. Within the framework of a novel mean field code based on Skyrme energy density functionals (Sky3D), the 1448 BNN-extrapolated $R_{\text{ch}}$ values yield a tiny standard deviation of 0.01 fm with respect to the Sky3D predictions, demonstrating the excellent $R_{\text{ch}}$ generalization capability of the BNN method~\cite{Chen2025,Chen2025_1}. To evaluate the accuracy of the improved method, the mean absolute $R_{\text{ch}}$ deviation, $\Delta R_{\text{ch}}$, between the improved method, and the BNN method is adopted, and $\Delta R_{\text{ch}}$ is given by
\begin{eqnarray}
	\Delta R_{\text{ch}}  =  \displaystyle \frac{1}{10}\sum^{10}_{i=1} |R_{\text{ch}}-R_{\text{ch}}^{\text{BNN}}| \, ,
	\label{eq:deltab}
\end{eqnarray}
where the $R_{\text{ch}}$ values from the BNN method are denoted as $R_{\text{ch}}^{\text{BNN}}$ for distinction. The summation runs over the ten nuclei. From Table~\ref{table:table0}, $\Delta R_{\text{ch}}$ is obtained as small as 0.01 fm, confirming the validity of the improved $\Delta B$-$R_{\text{ch}}$ relation for determining $R_{\text{ch}}$.

\begin{table}[tb]
	\caption{\label{table:table0}Comparison between $R_{\text{ch}}$ of ten neutron-deficient nuclei deduced using the present $\Delta B$-$R_{\text{ch}}$ relation by Eq.~(\ref{eq:eq14}), and the BNN-predicted values.}
	\centering
	\small
	\renewcommand{\arraystretch}{1.3}
	\begin{tabular*}{\linewidth}{@{\hspace{10mm}\extracolsep{\fill}}ccc@{\hspace{10mm}}}
		%\begin{tabular}{m{1.cm}<{\centering}m{2cm}<{\centering}m{2cm}<{\centering} } %
		\toprule[1.0pt]
		\midrule[0.25pt]
		Nucl.  & $R_{\text{ch}}$ (fm)\footnotemark[1] & $R_{\text{ch}}^{\text{BNN}}$ (fm) \\ 
		\midrule[0.25pt]
		$^{26}_{15}\text{P}$    &3.208~(35) &3.196~(28)      \\
		$^{27}_{16}\text{S}$    &3.281~(42) &3.250~(26)      \\
		$^{28}_{16}\text{S}$    &3.264~(28) &3.248~(20)      \\
		$^{41}_{22}\text{Ti}$   &3.574~(19) &3.577~(29)      \\
		$^{43}_{23}\text{V}$    &3.616~(20) &3.614~(37)      \\
		$^{45}_{24}\text{Cr}$   &3.656~(17) &3.666~(31)      \\
		$^{46}_{24}\text{Cr}$   &3.677~(11) &3.679~(23)      \\
		$^{47}_{25}\text{Mn}$   &3.701~(17) &3.712~(28)      \\
		$^{50}_{26}\text{Fe}$   &3.727~(12) &3.748~(15)      \\
		$^{53}_{28}\text{Ni}$   &3.756~(15) &3.744~(14)      \\
		\bottomrule[1.0pt]
	\end{tabular*}
	\footnotetext[1]{The errors are from the propagation of uncertainties in $\tilde{a}_{\text{C}}$, experimental $B$ and $R_{\text{ch}}$.}
\end{table}

To further show the improvement of the present method over the previous approach~\cite{Liu2026}, we also calculate $\Delta R_{\text{ch}}$ using the data in Table 2 of Ref.~\cite{Liu2026}. For the earlier method, $\Delta R_{\text{ch}}$ is found to be 0.02 fm. The significant decrease of $\Delta R_{\text{ch}}$ from 0.02 fm to 0.01 fm distinctly demonstrates the superiority of the improved method for determining $R_{\text{ch}}$. The improved method provides a more reliable means to determine the charge radii $R_{\text{ch}}$ of unmeasured short-lived nuclei using existing $B$ and $R_{\text{ch}}$ databases, and further facilitates systematic investigation of nuclear shell evolution through the $R_{\text{ch}}$ analysis.

\subsection{$R_{\text{ch}}$ determination for unmeasured\\short-lived nuclei}
\label{sec:Rch_determination}

\begin{table}[htpb]
	\centering
	\caption{\label{table:table1}$R_{\text{ch}}$ deduced from the available $B$ and $R_{\text{ch}}$ values in the mirror pairs ~\cite{mass2021_1,Zhang2023,Wang2023,Yu2024,Paul2021,Angeli2004,Wang2021_1,Pineda2021} using the improved $\Delta B$-$R_{\text{ch}}$ relation given by Eq.~(\ref{eq:eq14}) for the 59 nuclei with no measured $R_{\text{ch}}$ data.}
	\renewcommand{\arraystretch}{1.3}  
	\begin{tabular*}{\linewidth}{@{\hspace{10mm}\extracolsep{\fill}}cccc@{\hspace{10mm}}}
		\toprule[1.0pt]
		\midrule[0.25pt]
		Nucl.  &$R_{\text{ch}}$ (fm)   & Nucl.    &$R_{\text{ch}}$ (fm) \\
		\midrule[0.25pt]
		$^{8}$C                 &  2.973~(55)  & $^{36}$Cl               & 3.334~(24)  \\
		$^{9}$C                 &  2.691~(45)  & $^{39}$Sc               & 3.530~(18)  \\
		$^{10}$C                &  2.616~(31)  & $^{40}$Sc\footnotemark[1]  & 3.519~(13)  \\
		$^{11}$C                &  2.513~(27)  & $^{40}$Ti               & 3.580~(23)  \\
		$^{11}$N                &  3.003~(40)  & $^{41}$Ti               & 3.574~(19)  \\
		$^{13}$N                &  2.548~(14)  & $^{42}$Ti               & 3.588~(12)  \\
		$^{17}$N                &  2.882~(63)  & $^{43}$Ti               & 3.597~(15)  \\
		$^{11}$O                &  3.234~(60)  & $^{43}$V                & 3.616~(20)  \\
		$^{12}$O                &  3.002~(46)  & $^{44}$V                & 3.621~(12)  \\
		$^{14}$O                &  2.661~(24)  & $^{45}$V                & 3.631~(7)   \\
		$^{15}$O                &  2.683~(15)  & $^{44}$Cr               & 3.669~(21)  \\
		$^{20}$F                &  2.832~(65)  & $^{46}$Cr               & 3.677~(11)  \\
		$^{21}$F                &  2.889~(47)  & $^{46}$Mn               & 3.673~(22)  \\
		$^{23}$Al               &  3.088~(25)  & $^{47}$Mn               & 3.701~(17)  \\
		$^{24}$Al               &  3.084~(21)  & $^{48}$Fe               & 3.729~(20)  \\
		$^{25}$Al               &  3.086~(9)   & $^{50}$Fe               & 3.727~(12)  \\
		$^{24}$Si               &  3.130~(30)  & $^{51}$Fe               & 3.747~(12)  \\
		$^{25}$Si               &  3.132~(29)  & $^{52}$Co               & 3.743~(11)  \\
		$^{26}$Si               &  3.137~(16)  & $^{53}$Co               & 3.740~(8)   \\
		$^{27}$Si               &  3.110~(9)   & $^{55}$Co               & 3.692~(7)   \\
		$^{32}$Si\footnotemark[1]  &  3.181~(40)  & $^{52}$Ni               & 3.775~(19)  \\
		$^{26}$P                &  3.208~(35)  & $^{53}$Ni               & 3.765~(15)  \\
		$^{27}$P                &  3.190~(22)  & $^{58}$Zn               & 3.831~(10)  \\
		$^{29}$P                &  3.169~(9)   & $^{59}$Zn               & 3.848~(11)  \\
		$^{33}$P                &  3.222~(28)  & $^{60}$Ga               & 3.897~(12)  \\
		$^{27}$S                &  3.281~(42)  & $^{62}$Ge               & 3.958~(11)  \\
		$^{28}$S                &  3.264~(28)  & $^{63}$Ge               & 3.959~(18)  \\
		$^{30}$S                &  3.226~(15)  & $^{75}$Sr               & 4.250~(16)  \\
		$^{31}$S                &  3.233~(8)   &                         &             \\
		\bottomrule[1.0pt]
	\end{tabular*}
	\footnotetext[1]{With experimental values, i.e., $R_{\text{ch}}(^{32}\text{Si})=3.153(12)$~fm~\cite{Konig2024} and  $R_{\text{ch}}(^{40}\text{Sc})=3.514(25)$~fm~\cite{Konig2023}, but not used in the $\tilde{a}_{\text{C}}$ extraction in Fig.~\ref{fig:fig1}.}
\end{table}

Mirror pairs with two available measured $B$ values and one available measured $R_{\text{ch}}$ value are searched in the existing $B$~\cite{mass2021_1,Zhang2023,Wang2023,Yu2024,Paul2021} and $R_{\text{ch}}$~\cite{Angeli2004,Wang2021_1,Pineda2021} databases. Here the recently measured masses with the advanced B$\rho$-defined isochronous mass spectrometry at the experimental cooler storage ring (CSRe) in Lanzhou~\cite{Zhang2023,Wang2023,Yu2024} and with the TITAN multiple-reflection time-of-flight mass spectrometer at TRIUMF Isotope Separator and Accelerator (ISAC) facility in Vancouver~\cite{Paul2021} are also included. A total of 59 unmeasured nuclei with $Z\geq 6$ are identified, most of which correspond to neutron-deficient nuclei. The $R_{\text{ch}}$ values for the 59 nuclei with no measured $R_{\text{ch}}$ data are for the first step determined from the available $B$ and $R_{\text{ch}}$ values in the mirror pairs using the improved $\Delta B$-$R_{\text{ch}}$ relation of Eq.~(\ref{eq:eq14}) with the average $\tilde{a}_{\text{C}}$ extracted in Fig.~\ref{fig:fig1}. The results are listed in Table~\ref{table:table1}.

\begin{figure}[tpb]
	\centering
	\includegraphics[scale=0.4]{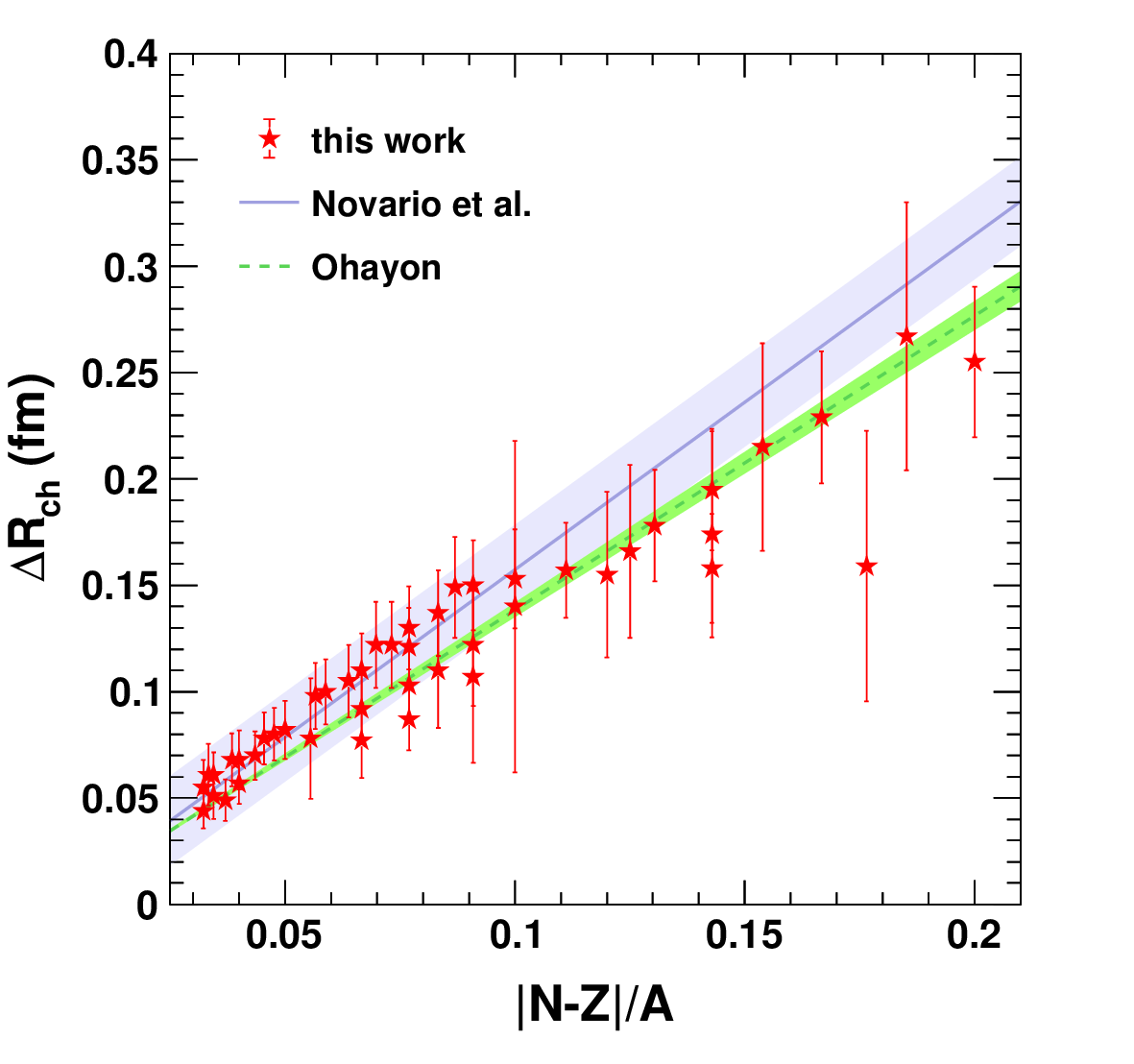}
	\caption{\footnotesize		
		Difference of the mirror-pair nuclear charge radii $\Delta R_{\text{ch}}$ versus $|N-Z|/A$. The data points are deduced from the $R_{\text{ch}}$ values listed in Table~\ref{table:table1} and the experimental values of the mirror partners. The lines with shaded bands are the linear regressions from the available experimental data, $\Delta R_{\text{ch}} = 1.381|N-Z|/A$ fm by Ohayon~\cite{Ohayon2025} (green band), and the coupled-cluster theory and the auxiliary field diffusion Monte Carlo method, $\Delta R_{\text{ch}} = 1.574|N-Z|/A$ fm by Novario \textit{et al}~\cite{Novario2023} (purple band), with a 1$\sigma$ confidence level.
	}
	\label{fig:fig3}
\end{figure}

Since the experimental $B$ and $R_{\text{ch}}$ values of $^{32}\text{Ar}/^{32}\text{Si}$ and $^{40}\text{Sc}/^{40}\text{K}$ mirror pairs are not incorporated in the current $\tilde{a}_{\text{C}}$ extraction in Fig.~\ref{fig:fig1}, the recently measured $R_{\text{ch}}$ values of $^{32}\text{Si}$~\cite{Konig2024} and $^{40}$Sc~\cite{Konig2023} can serve as an unbiased evaluation of the $R_{\text{ch}}$ determination performance using the present improved method. As listed in the table, $R_{\text{ch}}(^{32}\text{Si})$ and $R_{\text{ch}}(^{40}\text{Sc})$ are deduced to be 3.175~(39)~fm and 3.518~(13)~fm, respectively. These $R_{\text{ch}}(^{32}\text{Si})$ and $R_{\text{ch}}(^{40}\text{Sc})$ are in good agreement with those from the collinear laser spectroscopy experiments, i.e., 3.153~(12)~fm~\cite{Konig2024} and 3.514~(25)~fm~\cite{Konig2023}, respectively. The agreement verifies the validity of the present improved method for determining $R_{\text{ch}}$. It is also noted that the central value of $R_{\text{ch}}(^{32}\text{Si})$ deviates from the experimental value by 0.02 fm, a deviation larger than that is observed for $R_{\text{ch}}(^{40}\text{Sc})$. The origin of this deviation remains unclear at present. Clarifying it requires a dedicated future study, particularly as new data become available. Additional support for the present $R_{\text{ch}}$ determination can be obtained from the study of the difference of the mirror pair rms charge radii $\Delta R_{\text{ch}}$. In Fig.~\ref{fig:fig3}, the $\Delta R_{\text{ch}}$ versus $(Z-N)/A$ plot deduced from the $R_{\text{ch}}$ values in Table~\ref{table:table1} and the experimental values of the mirror partners are compared with the linear regression relationships from the available $\Delta R_{\text{ch}}$ data~\cite{Ohayon2025}, and the coupled-cluster theory and the auxiliary field diffusion Monte Carlo method~\cite{Novario2023}. As observed in the figure, our results show overall close agreement with both linear regressions in trend and magnitude, providing additional support for the present $R_{\text{ch}}$ determination.

\section{Evolution of nuclear shell structure in terms of $R_{\text{ch}}$}
\label{sec:evolution}

The promising performance of the improved $R_{\text{ch}}$ determination method permits us to systematically analyze the evolution of shell structure at $N=6$, 14, 20, and 28, using the obtained $R_{\text{ch}}$ values in Table~\ref{table:table1} and the corresponding experimental data~\cite{Angeli2004,Wang2021_1,Pineda2021,Wu2025,Konig2023}. To avoid the interference from pairing correlation effects when studying nuclear shell evolution, only the $R_{\text{ch}}$ values of the even-$N$ nuclei are adopted in the following discussion. 

\begin{figure}[tpb]
	\centering
	\includegraphics[scale=0.4]{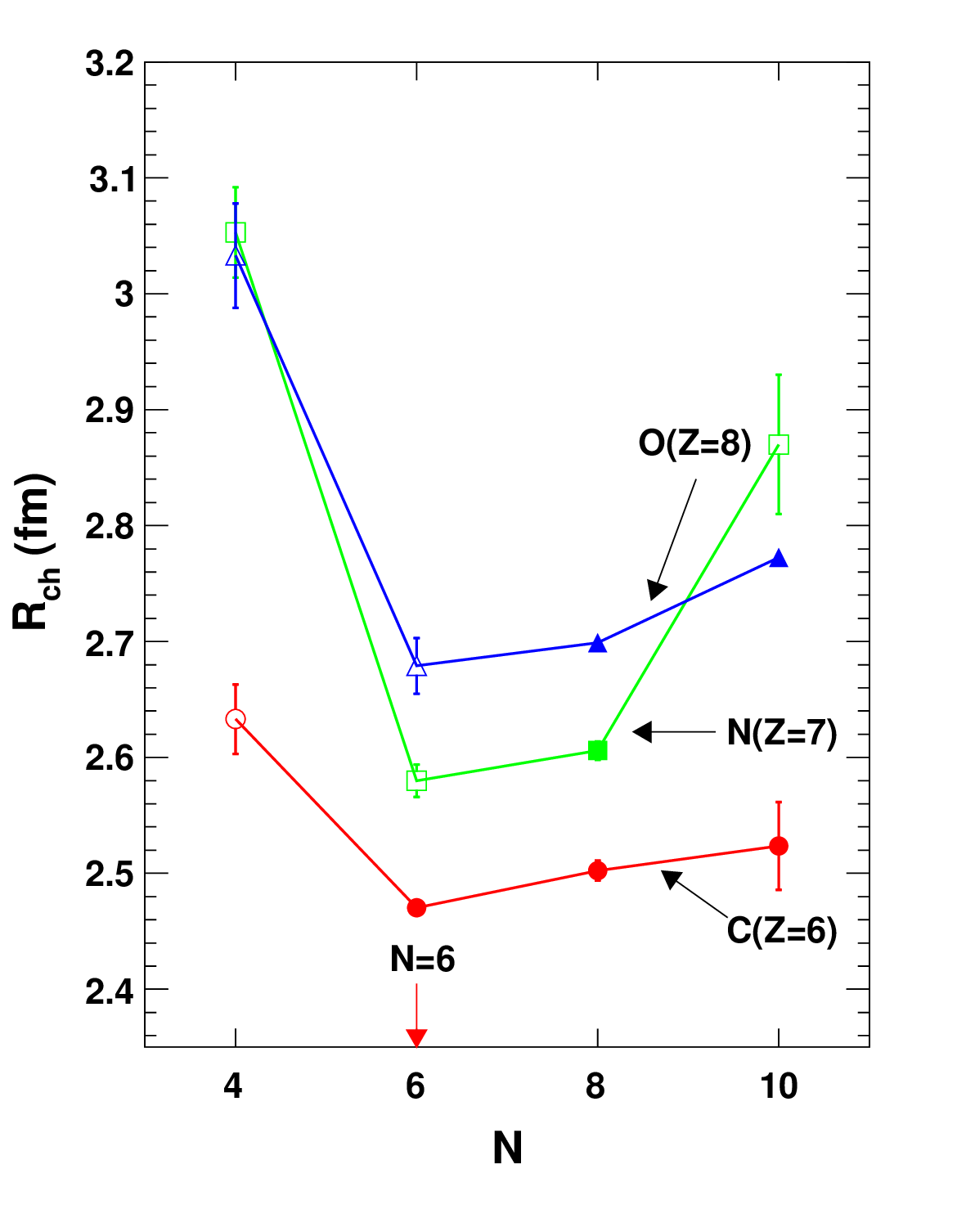}
	\caption{\footnotesize		
		Nuclear charge radii $R_{\text{ch}}$ as a function of $N$ for the carbon, nitrogen, and oxygen isotopes ($N=4, 6, 8, 10$). The newly obtained $R_{\text{ch}}$ values in this work (see $R_{\text{ch,avg}}$ in Table~\ref{table:table1}) are represented by open symbols, whereas the solid symbols are the experimental data taken from Refs.~\cite{Angeli2004,Wu2025}. The red arrow indicates the neutron number $N=6$. 
	}
	\label{fig:fig4}
\end{figure}

\subsection{$N=6$ subshell}

In Goeppert-Mayer's Nobel lecture, she mentioned the barely noticeable magic number 6, and surmised that the spin-orbit force creates a fairly small energy gap between the $1p_{1/2}$ and $1p_{3/2}$ orbits~\cite{Goeppert1963}. In 2001, the theoretical predictions of Otsuka \textit{et al.}~indicated that in neutron-rich exotic nuclei, a new $N=6$ magic number arises in the light mass region~\cite{Otsuka2001}. In 2018, Tran \textit{et al.}~found experimental evidence of a prevalent $Z = 6$ subshell closure in the neutron-rich C isotopes in terms of $R_{\text{ch}}$ trend variation based on the $R_{\text{ch}}$ measurements with charge-changing cross reactions~\cite{Tran2018}.

To pursue the evidence of the $N=6$ subshell closure from the perspective of $R_{\text{ch}}$, the newly obtained $R_{\text{ch}}$ values of the $^{10}\text{C}$, $^{11,13,17}\text{N}$ and $^{12,14}\text{O}$ isotopes are plotted together with the experimental data of measured isotopes in Fig.~\ref{fig:fig4}. With the newly obtained $R_{\text{ch}}$ values of the $\text{C}$, $\text{N}$ and $\text{O}$ isotopes, two distinct minima appear at $N=6$ and 8 along the three isotopic chains. Given the well-established $N=8$ shell closure in the present mass region, the minima at $N=6$ clearly indicate the significant $N=6$ subshell closure and the coexistence of $N=6$ and 8 magic numbers in both stable and neutron-deficient nuclei. Such findings are only possible to work out with the presently determined $R_{\text{ch}}$ data. It is noted that the recent theoretical calculations~\cite{Gupta2002,Lih2024} support these findings.

The $N=6$ subshell closure remains prevalent in the C, N, and O isotopic chains. This is in partial agreement with the results of \textit{ab initio} calculations by Li \textit{et al.}, that $N = 6$ is a local magic number in the $Z$ interval of $8\lesssim Z \lesssim 9$~\cite{Lih2024}. To confirm the range of persistence of the $N=6$ subshell closure, additional $R_{\text{ch}}$ data for more neutron-deficient and larger $Z$ nuclei, and corresponding systematic theoretical studies are required. One may also note that as $N$ continues to increase from 8 to 10, the increasing trend of $R_{\text{ch}}$ of the C and O isotopic chains is gentler than that of N. Similar phenomena can be also found by comparing the recent \textit{ab initio} studies on these C \cite{Tran2018}, N \cite{Bagchi2019}, and O \cite{Ren2025} isotopes. To date, the underlying mechanism remains unknown, and thus experimental efforts are required to confirm the theoretical results.

\subsection{$N=14$ subshell}

As a magic number candidate, 14 was also mentioned in Goeppert-Mayer's Nobel lecture~\cite{Goeppert1963}. Many experimental efforts devoted to neutron-rich O isotopes have indicated a possible subshell closure at $N=14$~\cite{Becheva2006,Belleguic2001,Thirolf2000}. For instance, the excitation energy of the $2^+_1$ state of $^{22}$O was measured to be 3199~(8)~keV, almost two times higher than that of adjacent $^{18,20}$O, indicating the presence of $N=14$ subshell closure~\cite{Belleguic2001}. The small $B(E2)$ value of 12~(8)~$e^2$fm$^2$ deduced from the inelastic scattering experiments supports the strengthening of the $N=14$ shell gap~\cite{Thirolf2000}. In 2016, Becheva \textit{et al.}~studied the $N=14$ shell gap by measuring the proton elastic and $2^+_1$ inelastic scattering angular distributions of $^{22}$O. The deformation parameter was deduced to be $\beta=0.26~(4)$, much smaller than that of $^{20}$O, demonstrating a pronounced $N=14$ shell closure~\cite{Becheva2006}.

\begin{figure}[t]
	\centering
	\includegraphics[scale=0.4]{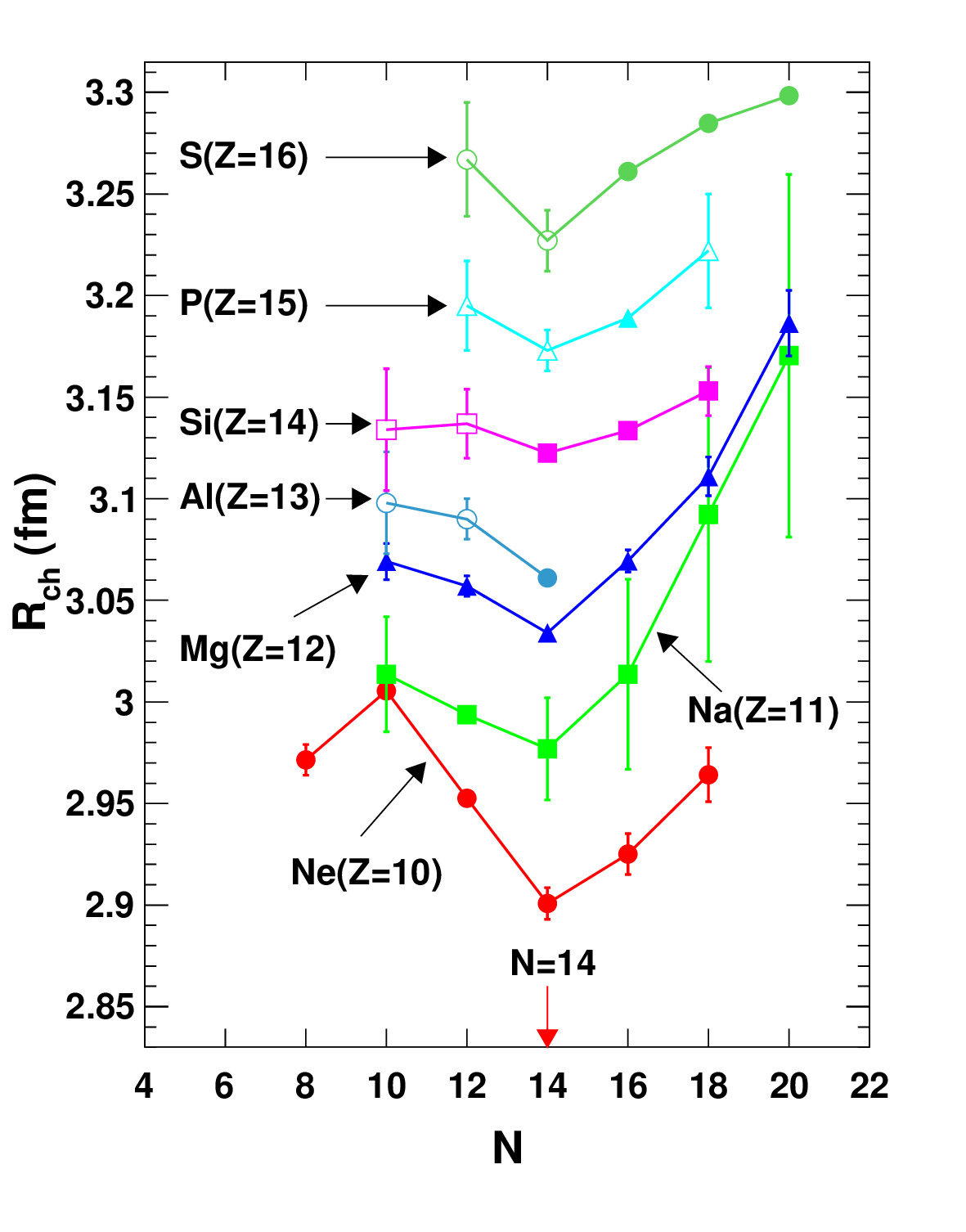}
	\caption{\footnotesize
		Nuclear charge radii $R_{\text{ch}}$ as a function of $N$ for the $Z=10$-16 isotopic chains. The red arrow indicates the neutron number $N=14$. The experimental data \cite{Angeli2004, Wang2021_1, Konig2024} (solid symbols) are supplemented by the newly obtained $R_{\text{ch}}$ values in this work (open symbols). 
	}
	\label{fig:fig5}
\end{figure}

The aforementioned studies merely focused the investigation of the $N=14$ subshell on differences in the isotopic properties of specific observables along a single isotopic chain in oxygen. Here we further analyze the $R_{\text{ch}}$ trend along various nearby isotopic chains, extending the investigation for the $N=14$ subshell closure to a broader $Z$ range. Similarly to Fig.~\ref{fig:fig4}, for $Z=10$-16 isotopic chains, the newly obtained $R_{\text{ch}}$ values for unmeasured isotopes are plotted together with the experimental data in Fig.~\ref{fig:fig5}. We observe that the $R_{\text{ch}}$ shows an identical kink with a minimum at $N=14$ as $N$ increases from 10 to 18, clearly demonstrating a strong $N=14$ subshell closure. The decrease of $R_{\text{ch}}$ at $N=8$ for $^{18}$Ne is attributed to the $N=8$ shell closure. The newly obtained $R_{\text{ch}}$ values of $^{23,25}$Al, $^{24,26}$Si, $^{27,29,33}$P and $^{28,30}$S extend the predominant range of the $N=14$ subshell closure from $Z=10$-12 up to $Z=16$. Recently, Bagchi \textit{et al.}~measured the $R_{\text{ch}}$ of neutron-rich $^{17-22}$N via charge changing reactions around 900~MeV/u at GSI~\cite{Bagchi2019}. A pronounced kink in $R_{\text{ch}}$ is also found at $N=14$, indicating that strong $N=14$ subshell closure also exists in the $Z=7$ isotopes. These evidences infer the existence of $N=14$ subshell closure. Future $R_{\text{ch}}$ measurements of neutron-rich O and F isotopes are desired to bridge the $Z$ gap between $Z=7$ and 10-16, supplementing more evidences of $N=14$ subshell closure.

Nevertheless, the $N=16$ subshell closure, identified in $^{24}$O from the large neutron subshell energy gap between the $2s_{1/2}$ and $2d_{3/2}$ orbits~\cite{Sorlin2008}, is not observed in Fig.~\ref{fig:fig5}. This may reflect the magicity of $N=16$ is somehow local, a feature observed exclusively in extremely neutron-rich nuclei, similar to the cases of $N=32$ and 34~\cite{Ye2025,Wienholtz2013,Steppenbeck2013,Koszor2021}. The perspective of attaining valuable insight into the $N=16$ subshell properties may also become accessible via further $R_{\text{ch}}$ measurements for neutron-rich O isotopes. Moreover, for the P and S isotopic chains, as $N$ decreases at $N<14$, a steep increase of $R_{\text{ch}}$ is observed compared to those of other nearby isotopic chains.

\subsection{$N=20$ shell}

The magic number 20 originates naturally from harmonic oscillator potential well descriptions of mean-field interactions in nuclear fermionic systems. The shell gaps at $N,Z = 20$ form between the $1d_{3/2}$ and $1f_{7/2}$ orbits. Nevertheless, one of the earliest evidences of shell closure reduction in the nuclear chart was reported for $N = 20$~\cite{Sorlin2008}, suggesting a new view to standard textbooks. Magic numbers are now known to evolve far from stability due to complex proton-neutron interactions. 

\begin{figure}[t]
	\centering
	\includegraphics[scale=0.4]{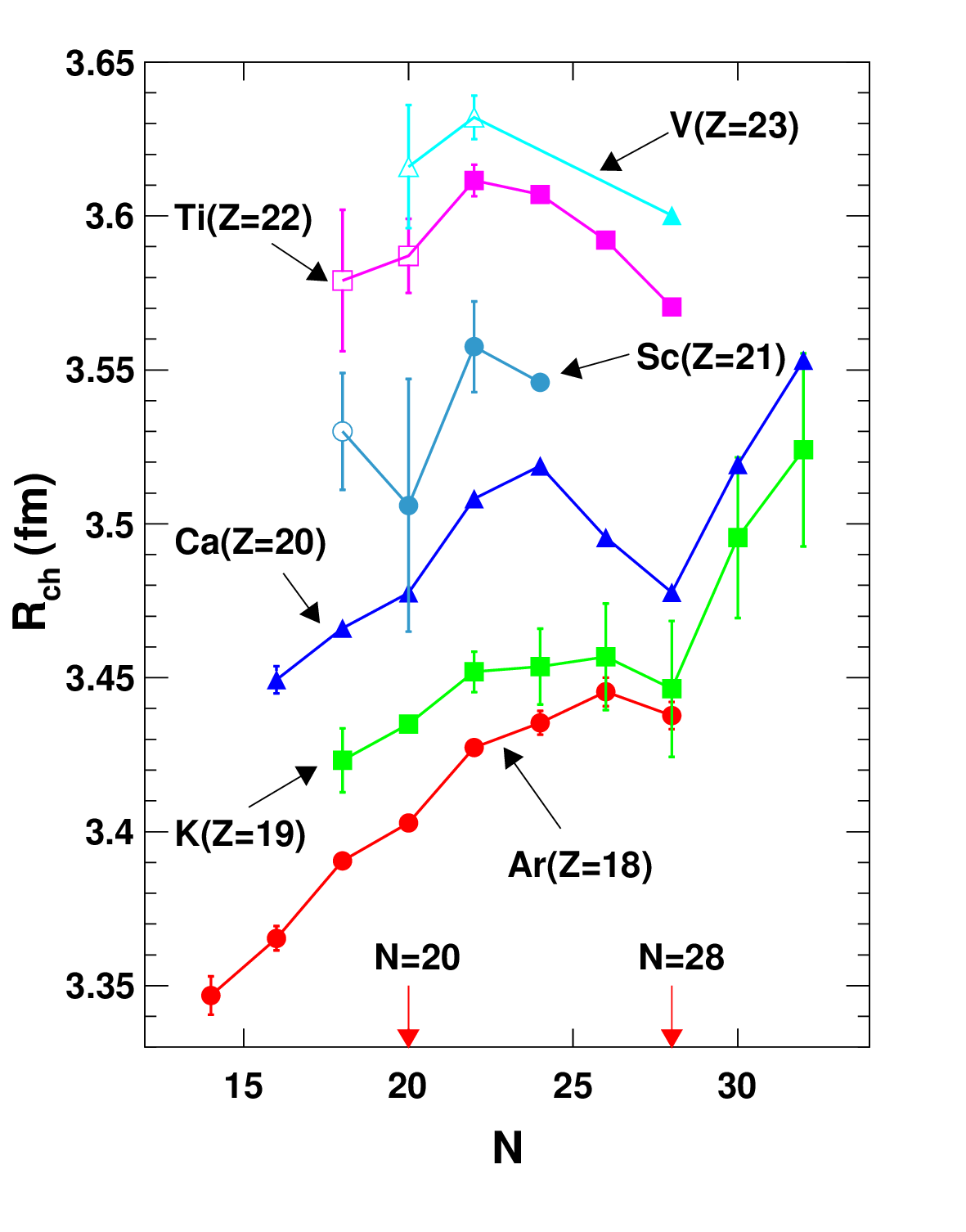}
	\caption{\footnotesize
		Nuclear charge radii $R_{\text{ch}}$ as a function of $N$ for the $Z=18$-23 isotopic chains. The red arrows indicate the neutron magic number $N=20$ and 28. The experimental data \cite{Angeli2004, Wang2021_1, Konig2023} (solid symbols) are supplemented by the newly obtained $R_{\text{ch}}$ values in this work (open symbols).
	}
	\label{fig:fig6}
\end{figure}

A recent example of the disappearance of the $N=20$ magic number is from that of the Ca isotopic chain. Only a smooth variation in $R_{\text{ch}}$ has been observed~\cite{Miller2019}. The absence of local minima is also found in the neighboring Ar~\cite{Klein1996} and K~\cite{Koszorus2021} isotopic chains. See the data summary in Fig.~\ref{fig:fig6}, where for $Z=18$-23 isotopic chains, the $R_{\text{ch}}$ values for unmeasured isotopes determined in this work are plotted together with the experimental values. The smooth variation trends of K and Ca isotopes near $N=20$ have  been well interpreted using \textit{ab initio} and density functional methods~\cite{Konig2023,Koszorus2021,Miller2019}.

More strikingly, with the newly obtained $R_{\text{ch}}$ of $^{39}$Sc, a pronounced kink at $N=20$ appears in the Sc isotopic chain, which can act as a promising signature of the $N=20$ shell closure. This result is of great interest and is well demonstrated by K\"{o}nig \textit{et al.}~\cite{Konig2023} based on the $R_{\text{ch}}$ measurements for $^{40,41}$Sc, i.e., $R_{\text{ch}}(^{40}\text{Sc})>R_{\text{ch}}(^{41}\text{Sc})$, using collinear laser spectroscopy. As a response to the proposal for  $R_{\text{ch}}$ measurements in the Ti isotopic chain across $N=20$~\cite{Konig2023},  the $R_{\text{ch}}$ values of $^{40,42}$Ti are determined. Interchain comparison of the determined $R_{\text{ch}}$ and the experimental value of $^{44}$Ti shows that the smooth decreasing trend with no discernible minimum appears again near $N = 20$ with the decrease of $N$ along the Ti isotopic chain. This finding definitely reveals the re-disappearance of the $N = 20$ shell closure when adding additional four proton outside the closed $1d_{3/2}$ shell. Such an abnormal ``disappearance $\rightarrow$ appearance $\rightarrow$ disappearance" evolution of the $N=20$ shell closure across $Z=20$-22 is observed for the first time, and may pose a significant puzzle for nuclear theory. Currently, the $R_{\text{ch}}$ determination for neutron-deficient V isotopes reaches $^{43}$V. To confirm whether the $N = 20$ shell closure disappears or not in the V isotopic chain, the $R_{\text{ch}}$ measurement for $^{41}$V is urgently required. As the $B$ and $R_{\text{ch}}$ values of its mirror partner $^{41}$Ar have been well known~\cite{Angeli2004,Wang2021_1}, the measurement for the $^{41}$V mass is also an alternative to mitigate the experiment difficulty. Then,  the $R_{\text{ch}}$ of $^{41}$V can be determined using the present improved method.

In addition to the puzzling ``disappearance $\rightarrow$ appearance $\rightarrow$ disappearance" evolution of the $N = 20$ shell closure, another notable feature observed in the figure is the presence of local maxima between $N =20$ and $N=28$ in the $Z=20$-23 isotopic chains. Shell model calculations of Ca indicate the local maxima are related to the partial breakdown of the $Z = 20$ shell closure caused by promotion due to the neutron-proton interaction~\cite{Caurier2001}. This characteristic trend becomes less pronounced in the K and Ar isotopic chains.  The feature difference at around $N=24$ between the K and Ca chains has been interpreted by the density functional theory and valence-space in-medium similarity renormalization group (depending on the choice of the functional or chiral nuclear force)~\cite{Koszorus2021,Miller2019,Baisw2025}. Nevertheless, the present $R_{\text{ch}}$ values of $^{40,42}$Ti combined with existing experimental Ti data show a maximum at $N=22$, two neutrons shift from $N=24$, and Sc chain displays the local maximum at the same position. As found by Bai \textit{et al.}~\cite{Baisw2025}, despite a considerable success in reproducing the trends of the K and Ca chains near $N=24$, the current nuclear models fail to describe the trend of Sc isotopes. The shift in the maximum positions between the $Z=20$ and $Z=21$-22 ($Z=23$ case is not confirmed due to the lack of $^{47}$V data) isotopic chains may pose additional challenge to nuclear theory.

Moreover, as $N$ continues to increase, the $R_{\text{ch}}$ trends of the  K and Ca isotopic chains show local minima at $N=28$, indicating a strong $N=28$ shell closure in K and Ca isotopes. More systematic analysis for the $N=28$ shell in the $Z=18$-23 region is inaccessible due to the lack of experimental data of neutron-rich nuclei in the involved $Z$ range in Fig.~\ref{fig:fig6}.

\subsection{$N=28$ shell}
The magic number 28 originates from the spin-orbit coupling in atomic nuclei. The studies of the evolution of the $N = 28$ magic shell in nuclei far from stability provide an ideal tool for probing the nuclear forces, such as the spin-orbit and tensor terms, in nuclei with reduced binding energies and/or depleted interior densities.

\begin{figure}[t]
	\centering
	\includegraphics[scale=0.4]{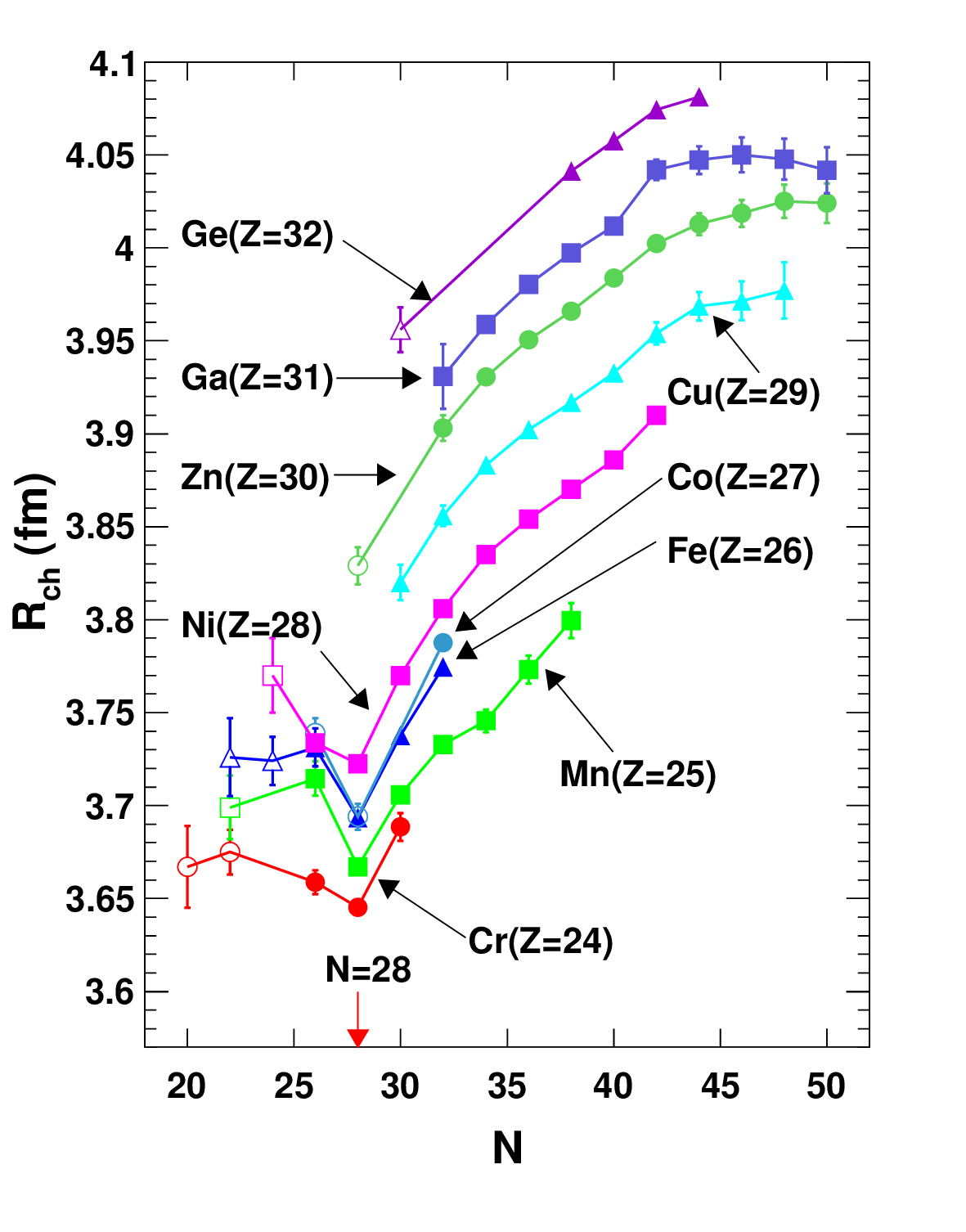}
	\caption{\footnotesize
		Nuclear charge radii $R_{\text{ch}}$ as a function of $N$ for the $Z=24$-32 isotopic chains. The red arrow indicates the neutron number $N=28$. 
		The experimental data \cite{Angeli2004, Wang2021_1, Pineda2021} (solid symbols) are supplemented by the newly obtained $R_{\text{ch}}$ values in this work (open symbols). 
	}
	\label{fig:fig7}
\end{figure}

Figure~\ref{fig:fig7} shows the newly obtained $R_{\text{ch}}$ and the experimental values taken from Refs.~\cite{Angeli2004,Wang2021_1,Pineda2021} for $Z=24$-32 isotopes. As observed, the $Z=24$-28 isotopic chains show appreciable minima at $N=28$ and the $R_{\text{ch}}$ values of the $Z=29$-32 chains decrease rapidly with $N$ approaching 28, indicating the robust closure of $N=28$ magic shell. At $N<28$, the $R_{\text{ch}}$ trend of the Ni isotopic chain and those of the Cr, Mn, and Fe chains diverge significantly. That is, as $N$ decreases at $N<28$, the $R_{\text{ch}}$ values of the Ni isotopes increase, whereas those of Cr, Mn, and Fe isotopes follow a parabolic trend beyond the $N=28$ kink similar to those of lower $Z$ isotopic chains as shown in Fig.~\ref{fig:fig6}. The underlying mechanism for such a difference is still unknown at present.

As also observed in Fig.~\ref{fig:fig7},  the $N= 32$ and 34 magicity which has been claimed by measuring the masses and the excitation energies of the $2^+_1$ states in neutron-rich isotopes around $^{52,54}$Ca~\cite{Wienholtz2013,Steppenbeck2013} is not clearly observed. This is possibly related to the local magic properties of $N= 32$ and 34, which are reflected by experimental results on nuclear masses and $2^+_1$ excitation energies to date~\cite{Ye2025}.

\section{Summary and perspectives}
\label{sec:summary}

In this article, the evolution of nuclear shell structure near $N=6$, 14, 20 and 28 (sub)shells is investigated using the probe of nuclear root-mean-square charge radii $R_{\text{ch}}$. To mitigate the issue due to the data scarcity for short-lived and exotic nuclei, we develop an improved method to determine unmeasured $R_{\text{ch}}$ values based on the correlation between binding energies $B$ and $R_{\text{ch}}$ of mirror nuclei. The improvement for this method consists of the consideration of the exchange term, charge-symmetry breaking effect and odd-even staggering effect in the Coulomb energy formulation, compared with that proposed in the Liu {\it et al.}~work~\cite{Liu2026}. Using the improved method, we determine the $R_{\text{ch}}$ values for 59 nuclei, most of which correspond to neutron-deficient nuclei, from their measured $B$ values and the corresponding  $B$ and $R_{\text{ch}}$ values of their mirror partners. With the newly obtained $R_{\text{ch}}$ values, we systematically analyze nuclear shell evolution near $N=6$, 14, 20 and 28 (sub)shells. The conclusions are summarized as follows:  
\begin{itemize}
	\item {
		\textit{$N=6$ subshell.} Two distinct minima of $R_{\text{ch}}$ appear at $N=6$ and 8 along the C, N and O isotopic chains, indicating the coexistence of $N=6$ and 8 magic numbers in both stable and neutron-deficient nuclei. The $N=6$ subshell closure persists across the studied $Z=6$-8 isotopic chains. As $N$ increases from 8 to 10, the $R_{\text{ch}}$ increasing trends for carbon and oxygen are softer than that of nitrogen.
	}
	\item {
		\textit{$N=14$ subshell.} With the newly obtained $R_{\text{ch}}$ values of $^{23,25}$Al, $^{24,26}$Si, $^{27,29,33}$P and $^{28,30}$S, the range of persistence for the $N=14$ subshell closure is extended from $Z=10$-12 to $Z=16$. No significant signal of the $N=16$ shell closure is observed across the studied $Z=10$-16 isotopic chains. For $N<14$, a more pronounced steepening trend of $R_{\text{ch}}$ in the P and S isotopic chains is evident.
	}
	\item {
		\textit{$N=20$ shell.} Abnormal ``disappearance $\rightarrow$ appearance $\rightarrow$ disappearance" evolution of the $N=20$ shell closure across $Z=20$-22 is observed, possibly posing a significant puzzle for nuclear theory. Between $N=20$ and 28, local maxima appear in the $N=20$-23 isotopic chains, but not in the 18-19 chains. These maxima positions differ between the $Z=20$ and $Z=21$-22 chains, i.e., at $N=24$ for $Z=20$, and at $N=22$ for $Z=21$-22. The $Z=23$ case remains unconfirmed due to the lack of $^{47}$V data.
	}
	\item {
		\textit{$N=28$ shell}. The $Z=24$-28 isotopic chains exhibit distinct minima at $N=28$, and for the $Z=29$-32 chains, the $R_{\text{ch}}$ values decrease rapidly as $N$ approaches 28, indicating the robust closure of $N=28$ magic shell. The $N= 32$ and 34 magicity is not observed in the studied isotopic chains. As $N$ decreases at $N<28$, the $R_{\text{ch}}$ values of the Ni isotopes increase, whereas those of Cr, Mn, and Fe isotopes follow a parabolic trend beyond the $N=28$ kink similar to those of lower $Z$ isotopic chains.  
	}
\end{itemize}

Most of the above insights into nuclear shell evolution are elucidated using the newly obtained $R_{\text{ch}}$ data. As a subsequent work, it will be essential to conduct systematic comparisons with the conclusions regarding nuclear shell evolution which have been drawn from other observables including nuclear masses, nuclear excitation energy gaps, magnetic dipole and electric quadrupole moments, and electromagnetic transition strength. Looking ahead, experimentally utilizing the improved method helps to alleviate practical experimental challenges, by converting the challenging $R_{\text{ch}}$ measurements for short-lived nuclei into more experimentally accessible $B$ measurements and the $R_{\text{ch}}$ measurements of their corresponding longer-lived mirror partners. This avenue holds considerable potential for nuclear shell evolution, proton-halo (or proton-skin) structure as well as other topics related to $R_{\text{ch}}$, as the unprecedented intensities of rare-isotope beams at  next-generation radioactive beam facilities~\cite{XHZhou2022,Brown2025,Geissel2003} enable the precise measurements of nuclear masses for a vast number of new and exotic nuclei far from the stability line. At present, the $R_{\text{ch}}$ measurements for exotic nuclei are challenging modern nuclear theories~\cite{Sun2025}.

The estimated $R_\mathrm{ch}$ values with the respective precision and study on the $N = 6, 14, 20$ and 28 (sub)-shell closure of this work serve as a forefront reference for future works based on the microphysics approaches, e.g., the charged radii based on the relativistic Hartree-Bogoliubov approach \cite{Liu2024} with PC-X \cite{Taninah2020}, PC-L3R \cite{Liu2023}, DD-PCX \cite{Yuksel2019}, and DD-MEX \cite{Taninah2020} interactions and relativistic Hartree-Bogoliubov approach in continuum \cite{Zhang2022,Guo2024} with PC-PK1 interaction \cite{Zhao2010}. Future synergy between expanded $R_{\text{ch}}$ data sets for the more short-lived nuclei and improved theoretical models will continue to deepen our understanding of nuclear shell evolution, unravel new nuclear physics phenomena in extreme isotopic regions, and advance the fundamental theoretical description of atomic nuclei.

\bigskip
\begin{acknowledgments}
	We greatly appreciate the anonymous Referee for meticulously reviewing our manuscript with useful suggestions and comments. We thank J. M. Dong (IMPCAS, Lanzhou) for fruitful discussions about the CSB calculations. This research was supported by National Natural Science Foundation of China (No. 12275186, No. 12322507, No. 12175156, and No. 11775277), Sichuan Science and Technology Program (No.2024NSFSC0420). YHL is appreciative of the Science Foundation of Zhejiang Sci-Tech University (No. 25062123-Y). 
\end{acknowledgments}
    
\bibliography{main_PRC}% Produces the bibliography via BibTeX.

\end{document}